\documentclass[sigconf]{acmart}

\renewcommand\footnotetextcopyrightpermission[1]{}
\AtBeginDocument{%
  \providecommand\BibTeX{{%
    \normalfont B\kern-0.5em{\scshape i\kern-0.25em b}\kern-0.8em\TeX}}}
    
\begin{document}
\title{Learning-To-Embed: Adopting Transformer based models for E-commerce Products Representation Learning}

\author{Lakshya Kumar}
\email{lakshyakri09@gmail.com}
\affiliation{%
 \institution{Myntra Designs Pvt. Ltd.}
 \country{India}}
 
\author{Sreekanth Vempati}
\email{sreekanth.vempati@myntra.com}
\affiliation{%
 \institution{Myntra Designs Pvt. Ltd.}
 \country{India}}
 

\setcopyright{none}






\begin{abstract}
Learning low-dimensional representation for large number of products present in an e-commerce catalogue plays a vital role as they are helpful in tasks like product ranking, product recommendation, finding similar products, modelling user-behaviour etc. Recently, a lot of tasks in the NLP field are getting tackled using the Transformer based models and these deep models are widely applicable in the industries setting to solve various problems. With this motivation, we apply transformer based model for learning contextual representation of products in an e-commerce setting. In this work, we propose a novel approach of pre-training transformer based model on a users generated sessions dataset obtained from a large fashion e-commerce platform to obtain latent product representation. Once pre-trained, we show that the low-dimension representation of the products can be obtained given the product attributes information as a textual sentence. We mainly pre-train \textbf{BERT}, \textbf{RoBERTa}, \textbf{ALBERT} and \textbf{XLNET} variants of transformer model and show a quantitative analysis of the products representation obtained from these models with respect to \textbf{N}ext \textbf{P}roduct \textbf{R}ecommendation(NPR) and \textbf{C}ontent \textbf{R}anking(CR) tasks. For both the tasks, we collect an evaluation data from the fashion e-commerce platform and observe that XLNET model outperform other variants with a Mean Reciprocal Rank(MRR) of \textbf{0.5} for NPR and Normalized Discounted Cumulative Gain(NDCG) of \textbf{0.634} for CR. XLNET model also outperforms the Word2Vec based non-transformer baseline on both the downstream tasks. To further measure the performance of the representations obtained from the transformer based model, we use the products representation obtained from our best model in one of the downstream ranking usecase and conduct an A/B test which showed significant improvements in different business metrics. To the best of our knowledge, this is the first and novel work for pre-training transformer based models using users generated sessions data containing products that are represented with rich attributes information for adoption in e-commerce setting. These models can be further fine-tuned in order to solve various downstream tasks in e-commerce, thereby eliminating the need to train a model from scratch.
\end{abstract}


\maketitle
\pagestyle{plain}

\section{Introduction}
In an e-commerce setting, latent product representation gives an edge to solve various tasks. These representations can be directly utilized in similarity based models and also serve as good latent features for training various machine learning and deep learning models for the downstream tasks. Different techniques are used in the past starting from traditional matrix factorization based approaches to neural network based approach for representing products in a low-dimensional space. At an abstract level, these embeddings(we use `representation' and `embeddings' interchangeably in this paper) tried to capture the notion that \textit{`similar things appear in similar context'}. In the recent past, the tremendous growth in Natural Language Processing gave novel models like transformer\citep{transformers} and its variants that are very effective in learning a deeper understanding of natural language. These models introduced the concept of self-attention which is used to capture the relationships between different tokens in a sentence, thereby learning the token representation that is dependent on the context. With the help of unsupervised tasks like Masked Language Modelling\cite{BERT} and Permutation Language Modelling\cite{xlnet}, a model can generate a lot of training examples in an unsupervised manner in order to perform the pre-training. The BERT model and its variants like RoBERTa\citep{Roberta} and ALBERT\citep{albert} showed a remarkable performance over standard benchmark datasets in GLUE\citep{GLUE}. The other models like XLNET\citep{xlnet} demostrated much better performance by introducing Permutation Language Modelling as the pre-training objective. These transformer variants can be directly pre-trained over the readily available users' sessions data in an e-commerce setting. This motivates us to pre-train these models over a users' sessions dataset collected from a top fashion e-commerce platform in order to obtain latent representation for various products. In a typical e-commerce setting, a single user session contains different product IDs\footnote{Each product id is a unique identifier of a product in an E-commerce Catalogue.} that can either be \textit{browsed, added to cart or purchased} by a user. In such user sessions, each product id can be replaced with the concatenation of various features/attributes of the product like description, price, style, pattern and other product specific attributes to create a textual sentence. By replacing every product id in a session with its textual sentence leads to a session paragraph which can be directly utilized for pre-training transformer variants. We leverage sessions paragraphs in order to pre-train the transformer based model and obtain contextual low-dimensional representation of various products. These low-dimensional product representations can be used in many downstream tasks like similar product retrieval, personalized product recommendations, ranking posts\footnote{A single post represent fashion content in the form of image, text and products(all inclusive).} for a user etc. Once pre-trained with the user-sessions data, transformer based model can be further fine tuned for any downstream task using the underlying data. The main contributions of our work are as follows:
\begin{enumerate}
    \item Introduce a novel approach for learning contextual low-dimensional products embedding from transformer based models.
    \item Train different transformer variants mainly BERT, RoBERTa, ALBERT and XLNET in an un-supervised manner on a users' sessions dataset in an e-commerce setting.
    \item Compare the performance of products embeddings obtained from pre-trained models with respect to Next Product Recommendation and Posts Ranking Task.
    \item Compare transformer based models with non-transformer based baseline, i.e., Word2Vec with respect to downstream tasks. 
    \item Highlight on an online experiment to show the improvements in business metrics with using transformer based products embedding. 
\end{enumerate}
The rest of the paper is organized as follows: In Section \ref{sec:related work}, we review previous literature works that highlight on learning the products representation. Section \ref{sec:approach} present our methodology for pre-training different transformer variants using a users' sessions dataset collected from a top fashion e-commerce platform. In Section \ref{sec:experimental setup}, we outline the experimental setup, implementation details and present the results of our evaluation with respect to Next Product Recommendation and Posts Ranking Tasks followed by discussions and a highlight on an online experiment. Finally, we conclude the paper and discuss future research directions in Section \ref{sec:conclusion}.

\section{Related Work}
\label{sec:related work}
Previous works on learning product representations in e-commerce propose very sparse models like Word2Vec\citep{mikolov:2013}. These models directly take the product id as input and try to bring the similar product closer in the latent vector space \citep{prod2vec}. There are research works that focus on using side information(in the form of product attributes like brand, pattern etc.) of the products and training Word2Vec models\citep{meta-prod2vec}. Other ways to learn the product representations is to use Collaborative Filtering (CF) to model users’ preferences based on their interaction histories \citep{Koren2011, 10.1145/371920.372071} that finally give latent representations of the products. Among various CF methods, Matrix Factorization (MF) is the most popular one, which projects users and items into a shared vector space and estimate a user’s preference on an item by the inner product between their vectors \citep{10.1109/MC.2009.263, 10.5555/2981562.2981720}. \citet{rendle2012bpr} show personalized ranking for individual users using pair-wise ranking loss minimization. They model the problem of matrix-factorization using Bayesian approach and optimize the pair-wise ranking loss with implicit user feedback. Recently, deep learning has been revolutionizing the representation learning dramatically. In the recent past with the development of Transformer\cite{transformers} model, researchers started adopting them for learning better contextualized representations. Transformer based models like BERT\cite{BERT} are also used recently for learning product representation using cloze task\cite{doi:10.1177/107769905303000401}. \citet{sun2019bert4rec} show how BERT model can be pre-trained in an unsupervised fashion in order to optimize for cloze objective and can be used for recommendation tasks(BERT4REC). They use item IDs in order to give input to the BERT model and show improvements over standard benchmark datasets for recommendations. There is another similar work like BERT4REC by \citet{BERTgoesshopping} where the authors directly apply the mask language modelling over user generated sessions in order to learn the product representations. While pre-training, \citet{BERTgoesshopping} apply the masking over product IDs in a user session in order to pre-train the model. Transformer model and its variants are primarily developed to learn contextual representation for solving different natural language tasks. These deep contextual model can understand way beyond than just the product IDs. In this work, we show how the transformer based model can directly take the textual data created from user generated sessions in order to learn the representation of individual tokens which can finally be used to represent the products. Representation obtained from such a model are based on rich textual attributes of the product. In the subsequent section, we explain our methodology of pre-training different transformer variants using a user generated sessions dataset to obtain low-dimension product representation. 

\begin{figure*}[h]
\centering
\includegraphics[scale = 0.15]{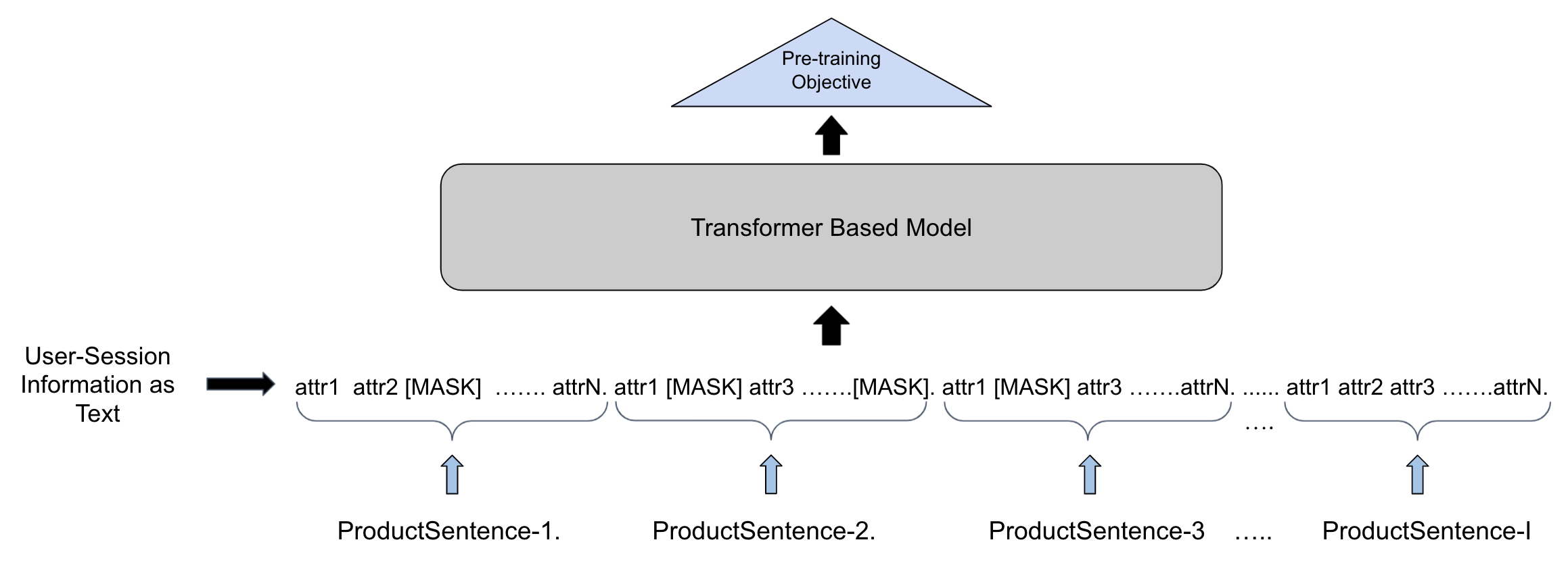}
\caption{Transformer Pre-training setup with User Session Data}
\label{fig:Training Setup}
\end{figure*}

\section{Approach}
\label{sec:approach}
In order to pre-train the transformer based models, we choose below auxiliary pre-training tasks: 
\begin{itemize}
     \item \textbf{Mask Language Modelling (MLM)}: Randomly masking some of the tokens in each sequence and predicting the masked tokens\citep{BERT}.
     \item \textbf{Permutation Language Modelling (PLM)}: Generating different permutations of the sequences and training a language model to learn better dependencies between the predicted tokens\citep{xlnet}.
 \end{itemize}
 
\begin{figure*}[h]
  \centering
  \includegraphics[scale = 0.20]{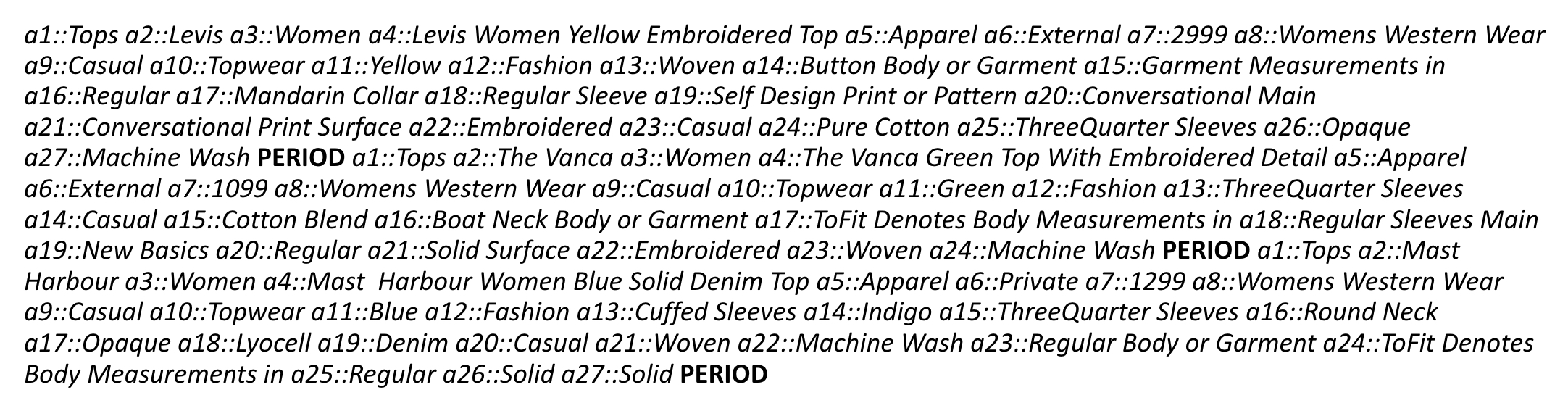}
  \caption{A user session paragraph}
  \label{fig:Dummy User Sentence}
\end{figure*}

We mainly pre-train BERT, RoBERTa, ALBERT and XLNET using a user generated sessions dataset. For BERT, RoBERTa and ALBERT, we simply optimize for Mask Language Modelling objective without taking into consideration the Next Sentence Prediction\cite{BERT} and the Sentence Order Prediction\cite{albert} objectives. For XLNET model, we pre-train using the Permutation Language Modelling objective. Masked Language Modelling objective has been shown as a sufficient task for pre-training by \citet{Roberta}, hence we only use MLM objective while pre-training BERT, RoBERTa and ALBERT. Figure \ref{fig:Training Setup} show our overall approach to pre-train different transformer models. The input is an example of a dummy user session\footnote{The actual user session is not included as it is confidential.} where one product sentence is formed by concatenating various product attributes. For example, ProductSentence-1 correspond to one product and is obtained by concatenating different attributes of the product as depicted in the Figure. Each of the product sentence are separated by full-stop\footnote{Ignore the white spaces between product sentences, they are just inserted to increase the readability.}. A dummy user session paragraph containing three product sentences constructed using product attributes is described in Section \ref{sec:experimental setup} and is shown in Figure \ref{fig:Dummy User Sentence}. In Figure \ref{fig:Training Setup}, the textual paragraph is composed of different product sentences which is tokenized using different tokenizer based upon the type of transformer variant used and then the embeddings corresponding to different tokens are used as input to the model. Note that in Figure \ref{fig:Training Setup}, [MASK] tokens are applicable only for BERT, RoBERTa and ALBERT model for MLM objective whereas XLNET model is pre-trained by predicting the next token given the previous tokens with the help of PLM objective. 
We use perplexity\citep{chen_beeferman_rosenfeld_2018} score for evaluating the pre-training performance of different models. Perplexity is defined as the exponentiated average log-likelihood of a sequence. For a tokenized sequence $X = (x_{0}, x_{1}, x_{2},...., x_{t} )$, perplexity is defined as, 
\begin{equation}
    PPL(X) = \exp{\{ \frac{-1}{t}\sum_{i}^{t}\log P_{\theta}(x_{i}|x_{<i})\}} 
\end{equation}
where $\log P_{\theta}(x_{i}|x_{<i})$ is the log-likelihood of the ith token conditioned on the preceding tokens $x_{<i}$ according to the model. After pre-training, the model would have learnt the syntactic and semantic aspects of product attribute tokens in a user session paragraph. With the help of self-attention it also try to learn the intra-product relation within attributes of a product and inter-product relation which exists between attributes of different products in a session. This helps in learning the contextual low-dimension representation of the products using the rich product attributes information.

\section{Experiments}
\label{sec:experimental setup}

\subsection{User session Data}
A user generated session consists of product IDs represented as: $Session_{u}: P_{id_{1}}, P_{id_{2}}, P_{id_{3}},..., P_{id_{n}}$. An example dummy user session\footnote{For confidential reasons the actual attribute names are replaced.} paragraph is shown in Figure \ref{fig:Dummy User Sentence}. Three different product sentences are separated by a period\footnote{For better readability, we have mentioned the word `PERIOD' instead of `.' after every product sentence.}. As the transformer based models are mainly developed to take textual data as the input, in order to generate a textual paragraph corresponding to a user's session, we replace every $P_{id_{i}}$ with a product sentence obtained by concatenating different attributes of the product: $P_{id_{i}}$. These attributes consists of high level attributes of the products like \textit{brand, product type, gender} etc and other product specific attributes which varies with different products. For every $P_{id_{i}}$ in a user session, we create these concatenated attribute sentence and again concatenate different sentences corresponding to different products to form a paragraph that represent a user session in the form of text. For pre-training different models, such paragraphs are given as input as shown in Figure \ref{fig:Training Setup} and once the  pre-training is completed, the latent product representation is obtained given the product attribute information in the form of textual sentence.

\subsection{Pre-training and Implementation}
We train the small versions of different transformer architectures for our experimentation. In each of the transformer architecture, the number of self-attention layers are kept as \textbf{6} and the number of attention heads per layer are kept as \textbf{12}. The input dimension of the self-attention layer is \textbf{768} and the size of the feed forward layer is \textbf{3072} across all transformer models. The \textbf{ALBERT} model use embedding factorization\citep{albert}, so the size of the embedding matrix is kept as \textbf{128} and the size of the projection matrix is \textbf{768}. In order to pre-train each of the model, we create a train and test dataset consisting of user sessions. As all of the transformer variants have an input limit of \textbf{512} tokens, we truncate user sessions to have only at most \textbf{20} products. We remove all the user sessions that contain less than 3 products. The number of sessions in the train and test data are \textbf{1 Million} and \textbf{0.5 Million} respectively. The tokenizers for different transformer models are trained on the train dataset with a vocabulary size of \textbf{30K}. And model to tokenizer mappings are as follows:
\begin{itemize}
    \item \textbf{BERT:} WordPiece\cite{wordpiece}
    \item \textbf{RoBERTa:} BPE\cite{BPE}
    \item \textbf{ALBERT/XLNET:} SentencePiece\cite{sentencepiece}
\end{itemize}
 The BERT, RoBERTa and ALBERT models are pre-trained to optimize MLM objective. While pre-training, the masking probability is kept as \textbf{0.15} which indicates that 15\% of the tokens in a user session paragraph are masked using the special [MASK] token as shown in Figure \ref{fig:Training Setup}. For XLNET model, we use the PLM objective for pre-training. Each of the transformer model is pre-trained using the \textbf{Pytorch}\citep{NEURIPS2019_9015} framework and the \textbf{HuggingFace} library\citep{wolf2019huggingfaces} based implementation. In order to do faster pre-training, the multi-gpu pre-training is done with \textbf{2 Tesla V100 GPUs} using Mixed Precision Training\citep{AMP_training}. The perplexity of different models after pre-training for \textbf{2 epochs} over the train and test data is shown in Table \ref{table:Perplexity}. Each of the model uses \textbf{AdamW} \citep{kingma2014method, loshchilov2017decoupled} optimizer in order to update the parameters. Once the models are pre-trained, we obtain the embedding of the products given the product attributes in the form of textual sentences from the last hidden layers of the models. The average of the tokens embedding from the last hidden layer is treated as product embedding. As we have pre-trained four different models, we obtain four different embeddings for each unique product in the dataset. In order to compare the transformer based models with the non-transformer approach, we also pre-train the Word2Vec\citep{word2vec} model using the Gensim\citep{gensim} library. The pre-training data that we use to train the Word2Vec model is same as that of the transformer based models. We train the Word2Vec model for 5 epochs with \textbf{768} as the hidden dimension and a minimum word frequency count of \textbf{20}. The vocabulory size of the Word2Vec model is \textbf{13109} words. The rest of the parameters of the Word2Vec model are used as default which are provided in the Gensim library. From the Word2Vec model, we obtain the embedding of the product by averaging the embeddings of the different words in the attribute based textual representation and finally use them for evaluation.
In order to compare different models, we take two downstream tasks: Next Product Recommendation and Posts\footnote{A single post represent fashion content having 3 things, i.e., image, textual description and products.} Ranking and discuss about the same in the next sub-sections.

\begin{table}
\centering
\begin{tabular}{ccc}
\hline \textbf{Model Type}  & \textbf{Train Perplexity} & \textbf{Test Perplexity}\\ 
\hline
BERT   & 1.260 & 1.247\\ 
RoBERTa & 1.282 & 1.267\\ 
ALBERT  & 1.129 & 1.130\\ 
XLNET  & 1.176 & 1.151\\ 
\hline
\end{tabular}
\caption{\label{table:Perplexity}Perplexity of Transformer variants }
\end{table}

\begin{table}
\centering
\begin{tabular}{ccc}
\hline \textbf{Model Type}  & \textbf{\#Parameters} & \textbf{MRR}  \\ 
\hline
Word2Vec & - & 0.43\\
BERT   & 66M & 0.49\\ 
RoBERTa & 66M & 0.48\\ 
ALBERT  & 56M & 0.47\\ 
XLNET  & 69M & \textbf{0.50}\\ 
\hline
\end{tabular}
\caption{\label{table:Next Item Recommendation Task}Next Product Recommendation Task}
\end{table}

\subsection{Downstream Task 1: Next Product Recommendation }
\label{sec:NPR}
After pre-training different models, we try to quantify the performance of learned representation of the products. The first task is Next Product Recommendation, where we predict the rank of the actual clicked product in a user session with respect to the randomly sampled negative\footnote{We call the clicked product as positive and the other randomly sampled products as negatives.} products and see how the learned representations are able to rank the actual clicked product given the user session in the form of vector representation. We create an evaluation data from user generated sessions from our e-commerce platform. This data consists of products browsed by a user and filtered as per the product category and gender. We filter the sessions as per the product category and gender in order to better represent them, as the user session may contain different categories of products even with different product gender. For example: a user session consists of some products, we first sort them with respect to their timestamps in a session and filter them with respect to their category and gender and consider those products that have same category and gender as one session. An example of category and gender can be \textit{dress} and \textit{women} respectively. The dataset consists of \textbf{0.47 Million sessions} and there are \textbf{0.7 Million} unique products in the dataset. For every product in the dataset, we generate the latent representation of the product from each of the pre-trained models. In order to generate the embedding, a product is represented by the concatenation of the attributes and after tokenization will be given as input to the transformer based models. The average of the tokens embedding from the last hidden layer of the model is treated as the product latent representation. For the Word2Vec model, we obtain the product representation by averaging the word embeddings of the words in the textual sentence(obtained by concatenating different attributes). For every product in the dataset, we obtain five sets of embeddings corresponding to four different pre-trained transformer variants and one non-transformer variant, i.e., Word2Vec. In order to evaluate these embeddings, we represent a user session in the form of embedding by taking the average of embedding of the first n-1 products which we call as session vector. The nth product in the session is combined with another set of randomly sampled negative products that are sampled from the same category and gender as of the nth product. The ratio of positive to negative is 1:20. We take the cosine similarity between the session vector and the 21 products and rank them based on their similarity score. Finally, this ranked list is used to compute the Mean Reciprocal Rank(MRR) across all user sessions. The MRR corresponding to different models is shown in Table \ref{table:Next Item Recommendation Task}. As we can see from the results, XLNET model outperform other models with a MRR of \textbf{0.50}. This explains that the representations that are learned from the XLNET model are able to better rank the actual clicked product in a user session as compared to other random products. The low-dimensional representation of a product encode different attributes information and co-relations between the attributes in the latent space. When we create a user session vector by averaging the n-1 product representations, the intent of a user in terms of different attributes also gets encoded in the latent space. Finally, a product whose attributes are very similar to the attributes information encoded in the session vector will get ranked higher as compared to other products. From Table \ref{table:Next Item Recommendation Task}, one can also see that almost all the transformer based models are having comparable number of parameters with XLNET model having the highest number of parameters, i.e., \textbf{69M}. It is also interesting to see that Word2Vec based non-transformer model having significantly less number of parameters is able to achieve an MRR of \textbf{0.43}.

\begin{table}
\centering
\begin{tabular}{llll}
\hline \textbf{Model Type} & \textbf{NDCG} \\ 
\hline
Word2Vec & 0.614 \\ 
BERT    & 0.631     \\
RoBERTa  & 0.629  \\ 
ALBERT   & 0.632 \\ 
XLNET   & \textbf{0.634} \\ 
\hline
\end{tabular}
\caption{\label{table:Posts Ranking Task}Posts Ranking Task}
\end{table}

\begin{figure*}[h]
  \centering
  \includegraphics[scale = 0.25]{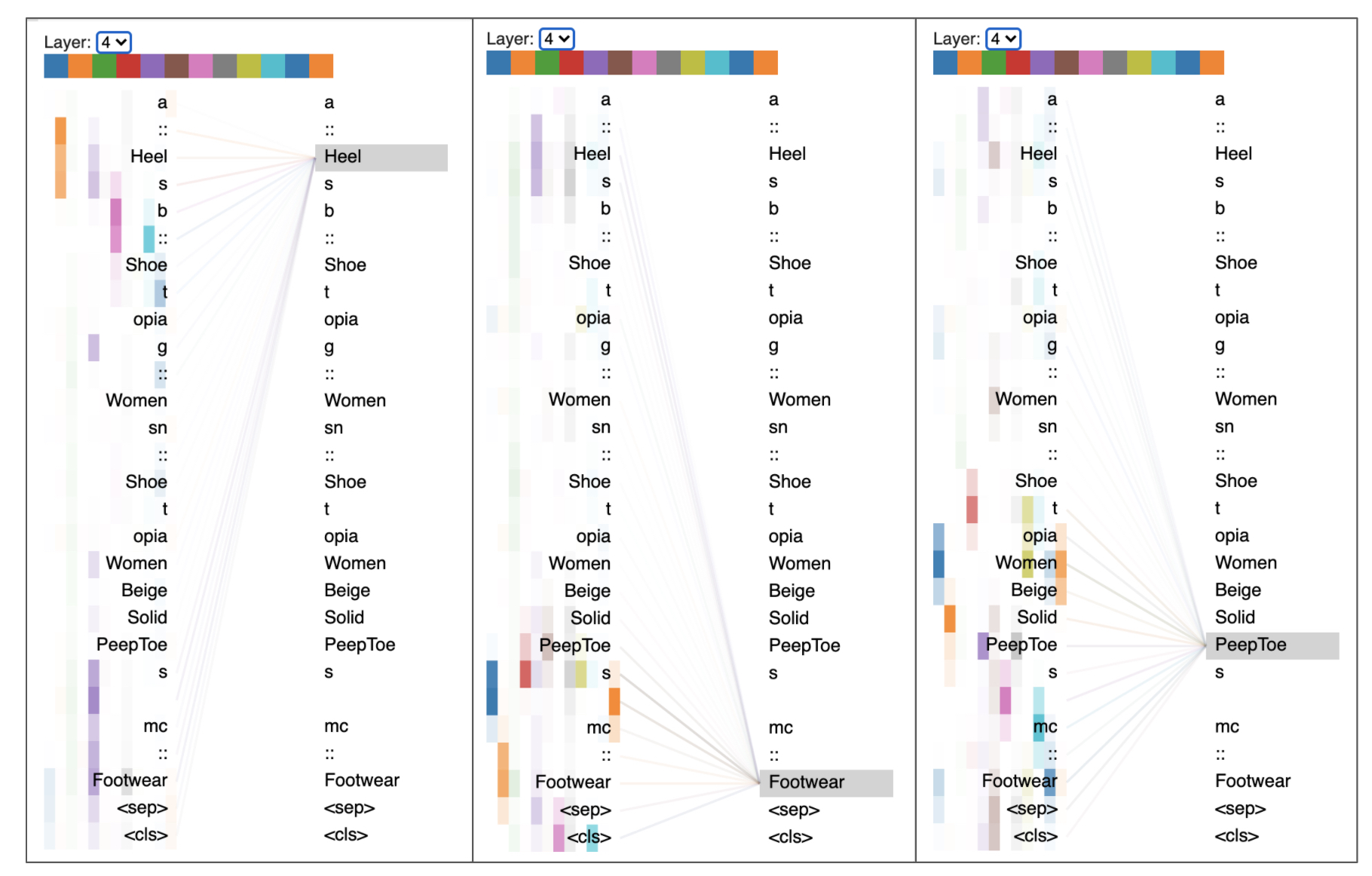}
  \caption{Visualization of the Attention Heads from the XLNET model for the product \textit{`Heel'}}
  \label{fig:Attention Heads}
\end{figure*}

\subsection{Downstream Task 2: Personalized Posts Ranking }

\subsubsection{\textbf{Dataset for Posts Ranking}}
\label{Dataset for Posts Ranking}
In order to further assess the performance of the learned products representation, we perform another downstream task of Posts Ranking. A posts dataset is taken from an in-house Social Commerce content based engagement feed\footnote{Feed indicates a ranked collection of posts.} within our e-commerce platform. An individual post in a feed is a content piece containing text and image showcasing different products by some fashion influencer. Each of the post contain different products that are explicitly tagged to the post by domain expert in an e-commerce setting. In order to generate the latent representation of a post, we average the representation of the products tagged in that post and call it as post latent representation. The number of unique posts in the dataset are \textbf{18K}. In the dataset, for every user there are posts with which user has engaged\footnote{In the dataset, engagement is defined as liking a post, sharing a post etc.} explicitly called as \textit{engaged posts} and there are posts which user just viewed\footnote{User spend some minimum time without engaging with a post.} called as \textit{viewed posts}. In order to extract features for a user, we prepare another dataset spanning a period of 1 year in which for every user all the engaged posts are extracted. 

\subsubsection{\textbf{Applying clustering for extracting User features}}
To extract users' features and finally represent them with top K posts, we run Ward\cite{Ward} Clustering algorithm that hierarchically clusters the posts using posts latent representation. For every user, we run the clustering over their last 1 year posts engagement data and obtain different clusters containing posts. Ward Clustering algorithm is hierarchical clustering, so it stops when the convergence criteria satisfies. We keep the distance threshold as the convergence criteria. After clustering converges, we find the medoid(post) from each cluster, whereby the medoid is the post having minimum euclidean distance with all the other post in the cluster. We obtain the medoids from each of the cluster and select the top K medoids in order to represent a user. For selecting the top K medoids, we use the timestamp of engagement of a user with the post and select those medoids that have higher engagement timestamp. For experiment, we fix K to be \textbf{10}, so for every user we extract at most top 10 posts after clustering. Every user is represented with the average of top 10 medoids(posts) embeddings. Note that the top 10 posts might vary across users as the engaged posts might be different in the dataset. Another important thing to note is that the Ward clustering does not learn any new latent representation. It just uses the already available representation to obtain the clusters in a bottom-up manner. 

\subsubsection{\textbf{Evaluation}}
As we have pre-trained five different models to obtain products representations, we obtain the posts representations from each of them as shown in the below mappings:
\begin{itemize}
    \item $Word2Vec \rightarrow Product_{Representation} \rightarrow Post_{Representation}$
    \item $BERT \rightarrow Product_{Representation} \rightarrow Post_{Representation}$
    \item $RoBERTa \rightarrow Product_{Representation} \rightarrow Post_{Representation}$
    \item $ALBERT \rightarrow Product_{Representation} \rightarrow Post_{Representation}$
    \item $XLNET \rightarrow Product_{Representation} \rightarrow Post_{Representation}$
    
\end{itemize}
A post is represented using products embeddings as mentioned in \ref{Dataset for Posts Ranking}. We run the clustering algorithm for every user five times(as we have five different models) with same hyperparameters and using five different posts representations. Thus, for every user, we obtain five latent representations(corresponding to five different models) by averaging the top 10 posts representation obtained after clustering. In the evaluation dataset, for every user, we have a set of engaged and viewed posts. In order to rank the viewed and the engaged posts for a user in the test data, we simply compute the cosine similarity between user embedding and post embedding. Finally, a ranked list is produced by sorting using the similarity score. We compute the Normalized Discounted Cumulative Gain(NDCG) metric in order to see the ranking performance of learned representations. The ground truth list for every user is taken as all the engaged posts followed by all the viewed posts. Table \ref{table:Posts Ranking Task} show the results with respect to different models. We observe that the XLNET model based products representation that are used to obtain the posts representation outperform with the NDCG of \textbf{0.634}. In the next subsection, we highlight over important points with respect to the pre-trained transformer variants and also explain the learned representation of a sample e-commerce product by visualizing the attention heads of our best performing model, i.e., XLNET. 

\subsection{Discussions}
\label{sec:discussions}
For each of the tasks, i.e., Next Product Recommendation and Posts Ranking, none of the transformer model is directly optimized. After pre-training each model over a user generated sessions dataset obtained from our e-commerce platform, we directly obtain the products embedding as explained in \ref{sec:NPR} and utilize them for the downstream tasks. Transformer based models are mainly developed for taking textual data as input in order to understand the syntactic and semantic aspects of the tokens based upon the other tokens in their context. As these deep contextualized language models offer a flexibility to take textual data as input rather than just taking the product unique identifiers, i.e., product IDs, we pre-train these models by giving user session paragraphs obtained by replacing product IDs with product attribute based sentences in order to make them learn the language of the e-commerce products. During pre-training, transformer based models try to understand the different attributes of the products as well as the context in which these attributes appear which further helps to develop a good understanding in terms of intra-product relation that exists between different attributes of a product as well as inter-product relation which exists between the attributes of different products. As per our experimentation, we see that the XLNET model outperforms the other three transformer variants on the two downstream task which is coherent with the fact that in the Natural Language for various tasks the Permutation Language Modelling objective(used in XLNET) is much better than the Mask Language Modelling as it helps in learning the dependencies between the predicted tokens in a much better way. We also observe that the XLNET model outperforms a non-transformer model, i.e., Word2Vec over both the downstream tasks. It is a novel approach of adopting the transformer based models in an e-commerce setting and going beyond product IDs and directly learning the product representations using the rich product attributes based textual sentences.\\ 
In Figure \ref{fig:Attention Heads}, we show visualization of the attention heads obtained from our pre-trained XLNET model in order to explain the learning of the model with respect to different product attributes. We have obtained these representations using BertViz\cite{bertviz} tool. For the visualization purpose, we choose 5th\footnote{Any other layer can be chosen, we have chosen the 5th layer (that is indexed as 4) as we see some clear explanability from it.} hidden layer of our model and show the learning of 12 different attention heads. If we look at the first visual, some of the attention heads learned the fact that keyword \textbf{`Heel'} is appearing in the context of keywords like \textbf{Shoetopia(brand)} and \textbf{Footwear(category)}. In the second visual, the keyword \textbf{`Footwear'} is appearing in the context of keywords like \textbf{Heel(product type)} and \textbf{Solid PeepToe(style of the product)}. Third visual is showing the learning of attention heads with respect to the keyword \textbf{PeepToe(style)}. The \textbf{`PeepToe'} is appearing in the context of keywords like \textbf{Heel(Product type), Shoetopia Women Beige Solid PeepToe(style)} and \textbf{Footwear(category)}. These visuals demonstrate that deep transformer based models are able to learn the relationships between different attributes of a product and try to bring those products closer in the latent space that share the similarities in terms of their attributes. In the next subsection, we discuss an online experiment that we conducted using the products embedding obtained from the transformer based model. 

\subsection{Online Experiment}
In order to see the real impact of the latent products representation learned from the transformer based model, we have done an online A/B testing for our Social Commerce Platform for Personalized User Feed(Posts) Ranking. We have taken the products embedding obtained from our best transformer model and used them to represent the posts as mentioned in \ref{Dataset for Posts Ranking}. Finally, we have used these posts representation in an already running pipeline to give personalized posts ranking for a user. We have compared this pipeline with an already deployed ranking solution in which Word2Vec based embeddings are getting used and observed $\sim$ \textbf{17\%} improvement in engagement per impression\footnote{Number of likes/clicks/shares are called as engagement} over posts and $\sim$ \textbf{10\%} improvement in impressions per user visit. Due to confidentiality constraints we are unable to disclose the full ranking pipeline that is getting used to obtain the personalized feeds for a user but these improvements in A/B test clearly indicates that the features that are learned using transformer based deep contextual models are significantly better as compared to shallow models like Word2Vec which can capture very limited context while learning the products latent representation.

\section{Conclusion \& Future Work}
\label{sec:conclusion}
We proposed a novel approach of learning contextual representation of e-commerce products using state-of-the-art transformer based models like BERT, RoBERTa, ALBERT and XLNET. Transformer based models are deep contextual models that can leverage rich textual information of the products in order to learn the correlations between different attributes and use that information with self-attention to learn the context of different attributes. With the help of unsupervised pre-training objectives like Masked Language Modelling and Permutation Language Modelling, many training examples can be generated in an unsupervised fashion with a user generated sessions dataset. We have pre-trained different Transformer variants mainly BERT, RoBERTa, ALBERT and XLNET in order to learn the latent products representation driven from rich product attributes. After pre-training, the products representation are directly obtained from the last hidden layers by averaging the tokens embedding. In order to compare the performance of latent representations from different models, we perform two task, i.e., Next Product Recommendation and Posts Ranking. In both of these tasks, the XLNET model outperforms the other three transformer variants. We also compare the transformer based models with a popular non-transformer baseline approach, i.e., Word2Vec. Our experiments indicate that XLNET model is able to outperform Word2Vec model on both the downstream tasks. We have also highlighted on an online A/B experiment which shows significant gains in the online business metrics for personalized posts ranking task. To the best of our knowledge, this is the first and the novel work that highlights the adoption of transformer based models in an e-commerce setting by pre-training using a user generated sessions dataset with rich product attributes.\\
As a future work, it will be interesting to directly fine-tune the transformer based models for the downstream task and compare their performance. Another direction is to pre-train other transformer architectures like GPT, GPT-2, Transformer-XL using a user sessions data and compare them with the ones which we trained in this research work. In this work, we directly take the token representation from the last hidden layers in order to generate product representation, it will be interesting to experiment with other layers in the transformer model and understand their learning in terms of syntax and semantics of the tokens. There are other interesting possibilities, for example taking other auxiliary tasks into consideration like Next Sentence Prediction and Sentence Order Prediction while pre-training in order to see their impact on the learned product representations. Also it will be interesting to incorporate other modalities like product images along with rich product attributes while learning products representation.

\bibliographystyle{ACM-Reference-Format}
\bibliography{sample-base}
\end{document}